\documentclass[twocolumn,english,aps,prl,twocolumn,showpacs,superscriptaddress]{revtex4}
\usepackage[T1]{fontenc}
\usepackage[latin9]{inputenc}
\usepackage{textcomp}
\usepackage{amsmath}
\usepackage{amssymb}
\usepackage{graphicx}
\usepackage{esint}

\makeatletter
\@ifundefined{textcolor}{}
{%
 \definecolor{BLACK}{gray}{0}
 \definecolor{WHITE}{gray}{1}
 \definecolor{RED}{rgb}{1,0,0}
 \definecolor{GREEN}{rgb}{0,1,0}
 \definecolor{BLUE}{rgb}{0,0,1}
 \definecolor{CYAN}{cmyk}{1,0,0,0}
 \definecolor{MAGENTA}{cmyk}{0,1,0,0}
 \definecolor{YELLOW}{cmyk}{0,0,1,0}
 }
\makeatother
\usepackage{babel}

\begin{document}

\title{Adiabatic State Conversion and Pulse Transmission in Optomechanical Systems}

\author{Lin Tian}

\email{LTian@ucmerced.edu}

\affiliation{University of California, Merced, 5200 North Lake Road, Merced, California 95343, USA}



\begin{abstract}
Optomechanical systems with strong coupling can be a powerful medium for quantum state engineering of the cavity modes. Here, we show that quantum state conversion between cavity modes of distinctively different wavelengths can be realized with high fidelity by adiabatically varying the effective optomechanical couplings. The conversion fidelity for gaussian states is derived by solving the Langevin equation in the adiabatic limit. Meanwhile, we also show that traveling photon pulses can be transmitted between different input and output channels with high fidelity and the output pulse can be engineered via the optomechanical couplings.
\end{abstract}
\maketitle

\emph{Introduction.} Light-matter interaction in optomechanical systems has been intensively explored \cite{reviews} and the strong coupling between the optical or microwave cavities and the mechanical modes was demonstrated in recent experiments \cite{strongcouplingExp1,strongcouplingExp2}. Electromagnetically induced transparency (EIT) and normal mode splittings have also been observed in such systems \cite{EIT1, EIT2, EIT3}.  It was shown that the mechanical modes can be prepared close to their quantum ground states in the resolved sideband regime \cite{cooling1, cooling2, cooling3, cooling4, Groundstate1, Groundstate2, Groundstate3, Groundstate4, Groundstate5}.

The optomechanical couplings can be explored for quantum state engineering of both the cavity and the mechanical modes. In earlier results, it was shown that sideband cooling can be realized on a mechanical mode by driving the cavity in the red sideband \cite{cooling1, cooling2, cooling3, cooling4}. It was also proposed that entanglement can be generated in an optomechanical system by driving the cavity in the blue sideband \cite{entanglement1}. The optomechanical systems have recently been studied as a medium for photon state transmission, storage, readout, and manipulation \cite{QND,control,strongcoupling1,pulsetransfer1, stateconversion1, pulsetransfer2, stateconversion3, pulsetransfer3, strongcoupling2, strongcoupling3}. In a previous work, we studied a scheme for quantum state conversion between cavity modes of distinctly different wavelengths by applying a sequence of $\pi/2$-pulses to swap the cavity and the mechanical states \cite{stateconversion1, stateconversion2}. The fidelity of this scheme is limited by cavity damping, thermal noise in the mechanical mode, and accuracy of the pump pulses. In particular, the fidelity shows a strong linear decrease with increasing thermal excitation number $n_{th}$.

Converting quantum states or traveling pulses between cavity modes with vastly different frequencies, such as an optical mode and a microwave mode, can have profound influence on quantum and classical information processing. In this work, we study the optomechanical system as a medium to transfer cavity states and to transmit photon pulses between different modes. Our result answers the outstanding question of how to overcome the effect of thermal noise on the transfer fidelity \cite{stateconversion1, stateconversion2}. We show that quantum states can be converted between different cavity modes by adiabatically varying the effective optomechanical couplings. During this process, the quantum states are preserved in a mechanical dark mode with negligible excitation to the mechanical mode. The concept of this scheme is similar to adiabatic state transfer in the EIT systems. The conversion fidelity for gaussian states shows negligible dependence on the thermal noise. Another advantage of this adiabatic scheme is that it does not require accurate control of the pump pulses. We also study the transmission of input pulses to a different output channel using this system. The condition for optimal transmission is derived in the frequency domain. High transmission fidelity can be achieved for input pulses with spectral width narrower than the relevant  transmission half-width. By applying time-dependent effective couplings, pulse engineering in the output channel can be realized. Our results indicate that quantum state transfer between vastly different input and output modes can be realized with high fidelity in this system. These results can facilitate the development of scalable quantum information processors containing photons, with applications in e.g. photon pulse generation and state manipulation, quantum repeaters, and conversion of information between optical and microwave photons \cite{opticsQIP}. 

\emph{Langevin equation in the adiabatic limit.} Our model is composed of two cavity modes and one mechanical mode coupling via optomechanical forces, which can be realized in various experimental systems \cite{Supplementary}. For cavity modes under external pumping, we follow the standard linearization procedure to derive the effective Hamiltonian for this coupled system \cite{cooling3, Supplementary, transformation},
\begin{equation}
H=\sum_{i=1,2}-\hbar\Delta_{i}a_{i}^{\dagger}a_{i}+\hbar g_{i}(a_{i}^{\dagger}b_{m}+b_{m}^{\dagger}a_{i})+\hbar\omega_{m}b_{m}^{\dagger}b_{m}\label{eq:Heff}
\end{equation}
where $a_{i}$ ($a_{i}^{\dag}$) is the annihilation (creation) operator for the $i$-th cavity mode ($i=1,2$), $b_{m}$ ($b_{m}^{\dagger}$) is for the mechanical mode, $\Delta_{i}$ is the laser detuning, $\omega_{m}$ is the mechanical frequency, and $g_{i}$ is the effective linear coupling that is proportional to the steady-state cavity amplitude \cite{entanglement1, stateconversion1}. To describe the system-bath coupling, we introduce the noise operators $a_{in}^{(i)}(t)$ for the $i$-th cavity mode and $b_{in}(t)$ for the mechanical mode. For simplicity of discussion, we choose the noise correlations $\langle a_{in}^{(i)}(t)a_{in}^{(i)\dagger}(t^{\prime})\rangle=\delta(t-t^{\prime})$ for the cavity modes and $\langle b_{in}(t)b_{in}^{\dagger}(t^{\prime})\rangle=(n_{th}+1)\delta(t-t^{\prime})$ for the mechanical mode at high temperature with the thermal excitation number $n_{th}$ \cite{Supplementary}. The cavity damping rates are $\kappa_{i}$ and the mechanical damping rate is $\gamma_{m}$. In our scheme, the pump laser is at the first red sideband with $-\Delta_{i}=\omega_{m}$ and the condition $|g_{i}|, \kappa_{i},\gamma_{m} \ll \omega_{m}$ is satisfied. Hence, the counter rotating terms $a_{i}^{\dagger}b_{m}^{\dagger}$ and $a_{i}b_{m}$ in the coupling, which generate a small heating on the mechanical mode as discussed in \cite{pulsetransfer2}, are neglected from the above Hamiltonian under the rotating wave approximation. The Langevin equation in the interaction picture can be written as \cite{Supplementary, QOtextbook}
\begin{equation}
id\vec{v}(t)/dt=M(t)\vec{v}(t)+i\sqrt{K}\vec{v}_{in}(t)\label{eq:Langevin}
\end{equation}
with the vector operators $\vec{v}(t)=[a_{1}(t),b_{m}(t),a_{2}(t)]^{\textrm{T}}$, $\vec{v}_{in}(t)=[a_{in}^{(1)}(t),b_{in}(t),a_{in}^{(2)}(t)]^{\textrm{T}}$, the dynamic matrix 
\begin{equation}
M(t)=\left(\begin{array}{ccc}
-i\frac{\kappa_{1}}{2} & g_{1}(t) & 0\\
g_{1}(t) & -i\frac{\gamma_{m}}{2} & g_{2}(t)\\
0 & g_{2}(t) & -i\frac{\kappa_{2}}{2}
\end{array}\right), \label{eq:Mt}
\end{equation}
and the diagonal matrix $K=\textrm{diag}(\kappa_{1},\gamma_{m},\kappa_{2})$. 

For time-dependent couplings $g_{i}(t)$, Eq.~(\ref{eq:Langevin}) can be solved under the adiabatic condition $|dg_{i}/dt|\ll g_{0}^{2}$ with $g_{0}=\sqrt{g_{1}^{2}(t)+g_{2}^{2}(t)}$ \cite{Supplementary, LandauZener}. Let $\lambda_{i}$ be the eigenvalues and $\psi_{i}$ be the eigenmodes of $M(t)$. For the transformation $U(t) = [\psi_{1},\psi_{2},\psi_{3}]$, we have $M(t)U(t) = U(t)\Lambda(t)$ with $\Lambda(t)=\textrm{diag}(\lambda_{1},\lambda_{2},\lambda_{3})$. In terms of the vector operators $\vec{\alpha}(t)=U^{-1}(t)\vec{v}(t)$ and $\vec{\beta}(t)=U^{-1}(t)\sqrt{K}\vec{v}_{in}(t)$, the Langevin equation can be transformed into
%
\begin{equation}
id\vec{\alpha}(t)/dt=i(dU^{-1}/dt)U(t)\vec{\alpha}(t)+\Lambda(t)\vec{\alpha}(t)+i\vec{\beta}(t).\label{eq:LangevinTransform}
\end{equation}
With $| [(dU^{-1}/dt)U(t)]_{ij} | \sim |dg_{i}/dt|/g_{0}\ll g_{0}$ \cite{Supplementary}, the first term on the right hand side of  Eq.~(\ref{eq:LangevinTransform}) can be neglected and the time evolution of the system operators can be derived as
\begin{equation}
\vec{\alpha}(t)=e^{-i\int_{0}^{t}dt^{\prime}\Lambda(t^{\prime})}\vec{\alpha}(0)+\int_{0}^{t}dt^{\prime}e^{-i\int_{t^{\prime}}^{t}dt^{\prime\prime}\Lambda(t^{\prime\prime})}\vec{\beta}(t^{\prime}).\label{eq:sol}
\end{equation}
Note that the operators used above are the shifted operators defined with regard to their steady-state amplitudes \cite{stateconversion1}. When the pump sources are adiabatically varied, the steady-state amplitudes follow the variation of the pump sources without affecting these equations.

\begin{figure}
\includegraphics[clip,width=7.5cm]{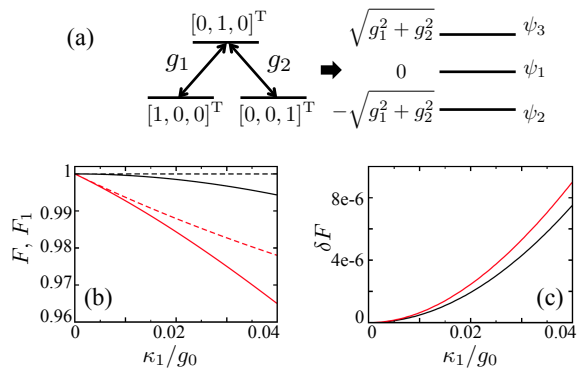}
\caption{(a) Mechanical dark mode. (b) $F$ (solid) and $F_{1}$ (dashed) for $|\alpha=1\rangle$ (upper curves) and $|\alpha=1,r=0.4\rangle$ (lower curves). (c) $\delta F$ for $|\alpha=1\rangle$ (lower curve) and $|\alpha=1,r=0.4\rangle$ (upper curve). Other parameters are $\kappa_{2}=0$, $g_{1}=5\sin(t)$, $g_{2}=-5\cos(t)$, and $T=\pi/2$ in arbitrary units.}
\label{fig1}
\end{figure}

\emph{Adiabatic cavity state conversion.} Under the two-photon resonance condition $\Delta_{1}=\Delta_{2}$ \cite{EITtheory1} and with $-\Delta_{i}=\omega_{m}$, quantum states can be converted between two cavity modes with high fidelity by adiabatically varying the couplings $g_{i}(t)$. The scheme is illustrated in Fig.~\ref{fig1} (a) for the simple case of zero dampings $\kappa_{i},\gamma_{m}=0$, where the eigenvalues of the matrix $M(t)$ are $\lambda_{1}=0$ and $\lambda_{2,3}=\mp g_{0}$ with an energy gap $g_{0}$ separating the modes. The eigenmode $\psi_{1}=[-g_{2},0,g_{1}]^{\text{T}}/g_{0}$ for $\lambda_{1}$ is a mechanical dark mode that only involves the cavity modes. The quantum state to be transferred is initially stored in mode $a_{1}$. The two other modes are in arbitrary single-particle states separable from mode $a_{1}$. At time $t=0$, $g_{2}$ starts at a large negative value and $g_{1}=0$, where the dark mode $\psi_{1}$ is simply the mode $a_{1}$ and $[\vec{\alpha}(0)]_{1}=a_{1}(0)$. Then, $-g_{2}(t)$ is adiabatically decreased to reach $g_{2}(T)=0$ at the final time $T$; and $g_{1}$ is adiabatically increased to reach a large positive value. The adiabatic condition requires that $T\gg 1/g_{0}$ in this scheme \cite{Supplementary}. At time $T$, the dark mode $\psi_{1}$ reaches the mode $a_{2}$ and $[\vec{\alpha}(T)]_{1}=a_{2}(T)$. During this whole process, the system is preserved in the mechanical dark mode. Using Eq.~(\ref{eq:sol}), we find that $a_{2}(T)=a_{1}(0)$, which shows that the initial state in mode $a_{1}$ has been transferred to mode $a_{2}$. In this scheme, the two-photon resonance condition is crucial for the existence of the mechanical dark mode which can be affected by the offset $\Delta_{1}-\Delta_{2}$ in the laser detunings \cite{Supplementary}.

This scheme is similar to adiabatic state transfer in the EIT systems where atoms in a $\Lambda$-system can be converted from one ground state to the other by adiabatically varying the Rabi frequencies \cite{EITtheory1, EITtheory2, LarsonPRA2005}. In our discussion, we let $-\Delta_{i}=\omega_{m}$. As we will show below, the mechanical noise has negligible effect on the state conversion in this regime. In comparison, in a Raman-like scheme with $\left|\Delta_{i}+\omega_{m}\right|\gg g_{i}$ \cite{quantumnetwork}, the state conversion can be realized via an effective Rabi flip with a Rabi frequency $\sim g_{1}g_{2}/|\Delta_{i}+\omega_{m}|$, where the cavity modes are prevented from mixing with the mechanical mode by the large energy offset $\left|\Delta_{i}+\omega_{m}\right|$ \cite{Supplementary}. 

For finite damping rates with $\kappa_{i},\gamma_{m}\ll g_{0}$, we treat the damping terms as perturbation \cite{Supplementary}. The eigenvalue of the mechanical dark mode becomes $\lambda_{1}=-i(\kappa_{2}g_{1}^{2}+\kappa_{1}g_{2}^{2})/2g_{0}^{2}$. The eigenvalues of the other eigenmodes are only slightly modified by the perturbation, and hence the adiabatic condition remains the unaffected. The mechanical dark mode becomes 
\begin{equation}
\psi_{1}=[-\frac{g_{2}}{g_{0}},-\frac{i(\kappa_{1}-\kappa_{2})g_{1}g_{2}}{2g_{0}^{3}},\frac{g_{1}}{g_{0}}]^{\text{T}},\label{eq:darkstate}
\end{equation}
which includes a small contribution from the mechanical mode and is not totally ``dark''. Using Eq.~(\ref{eq:sol}), we derive
\begin{equation}
a_{2}(T)=e^{-f(0,T)}a_{1}(0)+\int_{0}^{T}dt^{\prime}e^{-f(t^{\prime},T)}\beta_{1}(t^{\prime})\label{eq:a2}
\end{equation}
where $f(t,T)=i\int_{t}^{T}dt^{\prime}\lambda_{1}(t^{\prime})$ and $\beta_{1}(t)$ is composed of the noise operators in $\vec{v}_{in}(t)$ \cite{Supplementary}. With $\langle \vec{v}_{in}(t)\rangle=0$, we have $\langle a_{2}(T)\rangle=\exp[-\int_{0}^{T}dt^{\prime}(\kappa_{2}g_{1}^{2}+\kappa_{1}g_{2}^{2})/2g_{0}^{2}]\langle a_{1}(0)\rangle$, directly proportional to $\langle a_{1}(0)\rangle$ but with an exponential decay due to cavity damping. 

The fidelity of the state conversion can be defined as $F=(\textrm{Tr}[\sqrt{\sqrt{\rho_{i}}\rho_{f}\sqrt{\rho_{i}}}])^{2}$ on the final density matrix $\rho_{f}$ in cavity $a_{2}$ and the initial density matrix $\rho_{i}$ in cavity $a_{1}$. For gaussian states, the fidelity can be derived analytically once the covariance matrices of the initial and the final states are known \cite{gaussianfidelity}. Consider the initial state to be the squeezed state $|\alpha,\epsilon\rangle=D(\alpha)\exp((\epsilon^{\star}a_{1}^{2}-\epsilon(a_{1}^{\dagger})^{2})/2)|0\rangle$ where $D(\alpha)$ is the shift operator with amplitude $\alpha$ and $\epsilon=r\exp(2i\phi)$ with squeezing parameter $r$ \cite{QOtextbook}. At $r=0$, this state is the coherent state $|\alpha\rangle$. Using Eq.~(\ref{eq:a2}), the covariance matrix of the final state can be derived \cite{Supplementary}. The fidelity can be written as $F=F_{1}F_{2}$ with
\begin{subequations}
\begin{align}
F_{1} & \approx1-f(0,T)(\cosh(2r)-1)-f_{s}\cosh(2r)\label{eq:F1}\\
F_{2} & \approx1-f^{2}(0,T)y(\alpha,r)/2,\label{eq:F2}
\end{align}
\end{subequations}
where $f(0,T)\sim(\kappa_{1}+\kappa_{2})T/4$ linearly depends on the cavity damping rates and the term $f_{s}$ is due to the mechanical noise $b_{in}(t)$ with \cite{Supplementary}
\begin{equation}
f_{s}\lesssim\gamma_{m}(2n_{th}+1)T\left[(\kappa_{1}-\kappa_{2})/4g_{0}\right]^{2}. \label{eq:fs}
\end{equation}
When $\gamma_{m}(2n_{th}+1)\sim \kappa_{i}$, $f_{s}\ll f(0,T)$ and the factor $[(\kappa_{1}-\kappa_{2})/4g_{0}]^{2}$ significantly reduces the effect of the mechanical noise on the fidelity, which may be further reduced by engineering the damping rates to $\kappa_{1}\approx \kappa_{2}$. With $\kappa_{i},\gamma_{m}\ll g_{0}$, we expect the $f_{s}$ term can be much smaller than $f(0,T)$ even at room temperature \cite{strongcouplingExp1, strongcouplingExp2, EIT1, EIT2}. The function $y(\alpha,r)$ is composed of quadratic functions of $\alpha$ and $\alpha^{\star}$ and $y(\alpha,0)=2\left|\alpha\right|^{2}$ \cite{Supplementary}. The fidelity $F$ and $F_{1}$ are plotted in Fig.~\ref{fig1} (b). The factor $F_{1}$ decreases linearly with the cavity damping rates with $F_{1} \approx1-f_{s}$ for coherent states; the factor $F_{2}$, in contrast, decreases quadratically with the cavity damping rates. For illustration, we plot in Fig.~\ref{fig1} (c) the difference $\delta F$ between the conversion fidelities at $\gamma_{m}=0$ and at $\gamma_{m}/g_{0}=2\times 10^{-4}$ with $n_{th}=100$, which confirms that the mechanical noise has negligible effect on the fidelity.

In our previous work on quantum state conversion using $\pi/2$-pulses, the fidelity decreases with the mechanical noise as $-\gamma_{m}T(2n_{th}+1)\cosh(2r)/4$ \cite{stateconversion1}. In the current scheme, we exploit the mechanical dark mode which is immune to the mechanical noise to significantly reduce the effect of the mechanical noise. In addition, this adiabatic scheme does not require accurate control of the duration and magnitude of the pump pulses.

\begin{figure}
\includegraphics[clip,width=7.5cm]{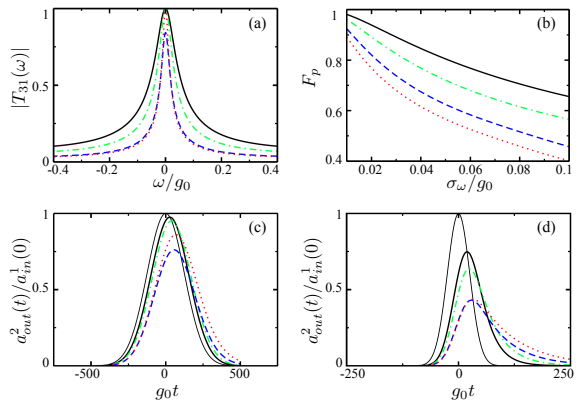}
\caption{(a) $|T_{31}|$, (b) $F_{p}$, and (c, d) $\langle  a_{out}^{(2)}(t) \rangle$ for $\sigma_{\omega}/g_{0}=0.008, 0.04$ respectively with $(\kappa_{1},\kappa_{2})/g_{0}$ being $(0.096,0.054)$ (solid), $(0.064,0.036)$ (dash-dotted), $(0.032,0.018)$ (dotted), and $(0.0192,0.032)$ (dashed). Other parameters are $\gamma_{m}/g_{0}=0.0002$, $g_{1}=4$, and $g_{2}=3$ in arbitrary units.}
\label{fig2}
\end{figure}

\emph{Pulse transmission and engineering.} Traveling photon pulses can be transmitted between input and output channels of distinctively different wavelengths. In our discussion, the input pulses have spectral width much narrower than the mechanical resonance. Consider a quantum input $a_{in}^{(1)}(t)$ in mode $a_{1}$, while $a_{in}^{(2)}(t)$ and $b_{in}(t)$ are noise operators with zero average. The output vector $\vec{v}_{out}(t)=[a_{out}^{(1)}(t), b_{out}(t), a_{out}^{(2)}(t)]^{T}$ can be derived using the Langevin equation and the input-output relation $\vec{v}_{out}(t)=\vec{v}_{in}(t)-\sqrt{K}\vec{v}(t)$ \cite{QOtextbook}. For constant effective couplings, the output pulse can be solved in the frequency domain with $\vec{v}_{in}(\omega)=\int (dt/\sqrt{2\pi})\vec{v}_{in}(t)e^{i\omega t}$ and $\vec{v}_{out}(\omega)=\int (dt/\sqrt{2\pi})\vec{v}_{out}(t)e^{i\omega t}$. We derive $\vec{v}_{out}(\omega)=\widehat{T}(\omega)\vec{v}_{in}(\omega)$ with the transmission matrix
\begin{equation}
\widehat{T}(\omega)=\left(I -i\sqrt{K}\left(I \omega-M\right)^{-1}\sqrt{K}\right)\label{eq:T}
\end{equation}
and the identity operator $I$. The transmission of the input pulse $a_{in}^{(1)}(\omega)$ to the output $a_{out}^{(2)}(\omega)$ is then characterized by the transmission matrix element $\widehat{T}_{31}(\omega)$ and the output pulse $a_{out}^{(2)}(t)$ can be calculated by integrating over the frequency components \cite{Supplementary}. In Fig.~\ref{fig2} (a), we plot the modulus $|\widehat{T}_{31}(\omega)|$ for four sets of damping rates $\kappa_{1,2}$ at $-\Delta_{i}=\omega_{m}$. The maximum of $|\widehat{T}_{31}(\omega)|$ occurs at $\omega=0$ which corresponds to the cavity resonances of modes $a_{1}$ and $a_{2}$. At the maximum, we have
\begin{equation}
\widehat{T}_{31}(0)=8g_{1}g_{2}\sqrt{\kappa_{1}\kappa_{2}}/(4g_{1}^{2}\kappa_{2}+4g_{2}^{2}\kappa_{1}+\gamma_{m}\kappa_{1} \kappa_{2}),
\label{eq:tr31}
\end{equation}
which gives the optimal transmission condition $g_{1}^{2}\kappa_{2}=g_{2}^{2}\kappa_{1}$ when $\kappa_{i}, \gamma_{m}\ll g_{i}$. Under this condition, $\widehat{T}_{31}(0)\approx 1$. It can also be shown that $\widehat{T}_{32}(\omega), \widehat{T}_{33}(\omega)\rightarrow 0$ as $\omega\rightarrow 0$, and hence the noise terms $a_{in}^{(2)}(\omega)$ and $b_{in}(\omega)$ are suppressed in the transmission. The transmission half-width $\Delta\omega$ defined by $|\widehat{T}_{31}(\Delta\omega)|=|\widehat{T}_{31}(0)|/2$ is
\begin{equation}
\Delta\omega\approx \sqrt{3}(g_{1}^{2}\kappa_{2}+g_{2}^{2}\kappa_{1}+\gamma_{m}\kappa_{1}\kappa_{2}/4)/2(g_{1}^{2}+g_{2}^{2}).\label{eq:halfwidth}
\end{equation}
These results indicate that a quantum input pulse $a_{in}^{(1)}(t)$ with a spectral width $\sigma_{\omega} \ll \Delta\omega$ can be transmitted with high fidelity to the output, while a pulse with $\sigma_{\omega} \gg \Delta\omega$ can be seriously deformed.
 
Below we study the photon transmission process by comparing the shapes of the input and output pulses. The pulse fidelity can be defined as \cite{Supplementary, pulsefidelity}
\begin{equation}
F_{p}=\frac{|\int dt \langle a_{in}^{(1)}(t)\rangle \langle a_{out}^{(2)}(t)\rangle^{\star}|^{2}}{\int dt | \langle a_{in}^{(1)}(t)\rangle|^{2}\int dt \langle a_{out}^{(2)}(t)\rangle|^{2}}.\label{eq:Fp}
\end{equation}
With the Cathy-Schwarz inequality, $F_{p}\le1$. The equality holds only at $\langle a_{in}^{(1)}(t)\rangle=c\langle a_{out}^{(2)}(t)\rangle$ which is equivalent to $\langle a_{in}^{(1)}(\omega)\rangle=c\langle a_{out}^{(2)}(\omega)\rangle$ with $c$ being a constant number. With $\langle a_{out}^{(2)}(\omega)\rangle=\widehat{T}_{31}(\omega)\langle a_{in}^{(1)}(\omega)\rangle$ for the frequency components, the pulse fidelity is thus determined by the properties of $\widehat{T}_{31}(\omega)$. Even though it does not fully quantify the transmission fidelity of quantum states, high pulse fidelity clearly indicates the possibility of high fidelity in the transmission of quantum states \cite{Supplementary}.

As an example, we study the transmission of an input pulse with the gaussian time-dependence $\langle a_{in}^{(1)}(t)\rangle=A \exp(-\sigma_{\omega}^{2}t^{2}/2)$ where $\sigma_{\omega}$ is the spectral width in the frequency domain. The normalization factor $A$ does not affect the pulse fidelity and we set $A=1$. The pulse fidelity decreases rapidly with the input spectral width as is plotted in Fig.~\ref{fig2} (b). For $(\kappa_{1},\kappa_{2})/g_{0}=(0.064,0.032)$, $\Delta\omega/g_{0}=0.04$. We have $F_{p}=0.97$ for $\sigma_{\omega}/g_{0}=0.008$ and $F_{p}=0.77$ for $\sigma_{\omega}/g_{0}=0.04$. For a given $\sigma_{\omega}$, the pulse fidelity is higher for larger transmission half-width. For $\sigma_{\omega} \ll \Delta\omega$, $\widehat{T}_{31}(\omega)\sim1$ in the entire spectral range of the input pulse so that $\langle a_{out}^{(2)}(t)\rangle \approx \langle a_{in}^{(1)}(t)\rangle $, giving high pulse fidelity. For $\sigma_{\omega} \gg \Delta\omega$, $\widehat{T}_{31}(\omega)$ decreases rapidly when $|\omega|>\Delta\omega$ and the output pulse is seriously deformed. In Fig.~\ref{fig2} (c, d),  we plot $\langle  a_{out}^{(2)}(t) \rangle$ for $\sigma_{\omega}/g_{0}=0.008,0.04$ to demonstrate the above analysis.  

Meanwhile, the output pulse can be engineered by applying time-dependent effective couplings. Using Eq.~(\ref{eq:sol}) and the relation $\vec{v}(t)=U(t)\vec{\alpha}(t)$, the output vector $\vec{v}_{out}(t)$ can be derived as an integral function of the input operator $a_{in}^{(1)}(t^{\prime})$ and the noise operators $a_{in}^{(2)}(t^{\prime})$ and $b_{in}(t^{\prime})$ during time $0\le t^{\prime} \le t$. The effective couplings $g_{i}(t)$ modulate the dependence of the output operator on the input operator and can hence manipulate the output pulse $a_{out}^{(2)}(t)$. This is presented in more detail in the Supplementary Materials \cite{Supplementary}. 

\emph{Conclusions.} We showed that quantum state conversion between modes with vastly different frequencies such as optical and microwave modes can be realized with high fidelity by an adiabatic scheme via the mechanical dark mode. The scheme is immune to the mechanical noise and does not require accurate control of the pump pulses. We also illustrated that high-fidelity transmission of quantum pulses between different input-output channels and pulse engineering in the output channel can be realized via the optomechanical couplings. Our work demonstrates that the optomechanical systems can be explored for photon state engineering and for various applications in quantum information processing. 

\emph{Acknowledgements.} This work is supported by the DARPA ORCHID program through AFOSR, NSF-DMR-0956064, NSF-CCF-0916303, and NSF-COINS. When finishing this project, we found the preprint arXiv:1110.5074 by Y.-D. Wang and A. A. Clerk on related subject.

\end{document}